\documentclass[fleqn,usenatbib]{mnras}
\usepackage{amsmath}
\usepackage{mathptmx}
\usepackage{txfonts}
\usepackage[T1]{fontenc}
\usepackage{ae,aecompl}
\usepackage{graphicx}
\usepackage{subfigure}
\usepackage{xcolor}
\usepackage{comment}
\usepackage{multirow}
\usepackage[normalem]{ulem}

\title[Dust accretion in binary systems]{Dust accretion in binary systems: implications for planets and transition discs}

\author[Y. Chachan et al.]{
Yayaati Chachan,$^{1,2,3}$\thanks{E-mail: ychachan@caltech.edu}
Richard A. Booth,$^{2}$
Amaury H. M. J. Triaud,$^{2,4}$
Cathie Clarke$^{2}$
\\
$^{1}$Division of Geological and Planetary Sciences, California Institute of Technology, Pasadena, CA, USA\\
$^{2}$Institute of Astronomy, University of Cambridge, Madingley Road, Cambridge CB3 0HA, UK\\
$^{3}$St John's College, Cambridge, CB2 1TP, UK\\
$^{4}$School of Physics \& Astronomy, University of Birmingham, Edgbaston, Birmingham B15 2TT, UK
}

\date{Accepted 2019 August 28. Received 2019 August 27; in original form 2018 July 31}

\pubyear{2019}

\begin{document}
\label{firstpage}
\pagerange{\pageref{firstpage}--\pageref{lastpage}}
\maketitle

\begin{abstract}
The presence of planets in binary systems poses interesting problems for planet formation theories, both in cases where planets must have formed in very compact discs around the individual stars and where they are located near the edge of the stable circumbinary region, where {\it in situ} formation is challenging. Dust dynamics is expected to play an important role in such systems, since dust trapping at the inner edge of circumbinary discs could aid {\it in situ} formation, but would simultaneously starve the circumstellar discs of the solid material needed to form planets. Here we investigate the dynamics of dust in binary systems using Smooth Particle Hydrodynamics. We find that all our simulations tend towards dust trapping in the circumbinary disc, but the timescale on which trapping begins depends on binary mass ratio (q) and eccentricity as well as the angular momentum of the infalling material. For $q \gtrsim 0.1$, we find that dust can initially accrete onto the circumstellar discs, but as the circumbinary cavity grows in radius, dust eventually becomes trapped in the circumbinary disc. For $q = 0.01$, we find that increasing the binary eccentricity increases the time required for dust trapping to begin. However, even this longer timescale is likely to be shorter than the planet formation timescale in the inner disc and is insufficient to explain the observed pre-transitional discs. This indicates that increase in companion eccentricity alone is not enough to allow significant transfer of solids from the outer to the inner disc.
\end{abstract}


\section{Introduction}
\label{sec:intro}

The discovery of circumbinary planets has required the extension of planet formation theories to a new dynamical regime and presented numerous challenges to the existing theoretical picture. One feature is particularly notable, i.e. that circumbinary planets tend to orbit just beyond the dynamical stability limits of the binary stars \citep{Welsh2012,Martin2014, Martin2015}. The semi-major axes of these planets is a few 10s of percent larger than the minimum distance of a long-term stable planet from the binary's centre of mass \cite{Holman1999}. This observation has led to a debate about the origin of such a dynamical arrangement and posed an interesting question for planet formation theories. Do the circumbinary planets preferentially form near their observed locations or do they form elsewhere and tend to migrate to their observed locations?

The preferential formation of circumbinary planets at their observed locations (i.e. {\it in situ}) is extremely challenging due to their close proximity to the binary stars. The dynamical influence of the binary in this region is strong and tends to pump up the eccentricities of planet forming material \citep{Moriwaki2004, Paardekooper2012, Meschiari2012a}. These large eccentricities result in high velocity collisions between planetesimals \citep{Marzari2000, Scholl2007}, causing their catastrophic disruption. This prohibits growth to the 100 km size range, which is more resilient to catastrophic collisions. Although apsidal alignment and slight mutual inclination between the binary and the disc can reduce the severity of this problem by reducing relative collision velocities, {\it in situ} formation of circumbinary planets remains somewhat problematical \citep{Marzari2000, Xie2009, Lines2014, Bromley2015}.

The possibility of {\it ex situ} formation and subsequent migration of circumbinary planets to their observed locations has received more support. Further out in the circumbinary disc \citep[i.e. well beyond the circumbinary cavity so typically at $>$ 10 AU for binary semi-major axis 1 AU;][]{Moriwaki2004, Scholl2007, Rafikov2013a}, the dynamical environment would seem as hospitable for planet formation as it is in discs around single stars. However, one issue faced by planet formation at large distances from the binary is the prohibitively long formation timescale associated with the longer orbital timescales at such distances. Intermediate regions are either subject to strong perturbations by the binary \citep{Moriwaki2004} or turbulence \citep{Meschiari2012b}. Inclusion of disc self-gravity helps to shift the inner boundary of the region where planetesimal accretion is viable slightly further in (though not to the current locations of the planets in observed systems around $\sim$ AU scale binaries), as it dampens the growth of planetesimal eccentricity and lowers collisional velocities \citep{Rafikov2013a, Silsbee2015}.

Even if planetary cores can form in the outer disc, there is still the challenge of migrating them and piling them up just beyond the stability limit. Migration in a circumbinary disc could happen via the standard Type I or II mechanisms but it needs to halt when planetary cores reach the observed planetary locations \citep{Nelson2003, Pierens2007, Meschiari2012a, Silsbee2015}. \cite{Pierens2007} suggested that the pressure maximum created beyond the inner edge of the truncated circumbinary disc could slow down or halt the inward migration of a planetary core \citep[see also][]{Masset2006}. While current simulations do not yet reproduce the observed location of planets \citep{Pierens2013, Kley2014, Mutter2017}, additional physics that is typically neglected such as radiative transport, magnetic fields and disc self-gravity may help to alleviate the discrepancy. Thus formation at larger distances, followed by inward migration that stalls near the cavity edge remains a possible explanation for the origin of circumbinary planets.

{\it In situ} growth of planetary embryos may turn out to be important, if dust can be effectively trapped in the pressure maximum at the inner edge of the circumbinary disc. First, if sufficient density of dust can build up, the dust might collapse directly to form planetary seeds surpassing the metre-sized barrier, perhaps further aided by the streaming instability \citep{Youdin2005, Johansen2007}. Second, a high density of dust would lead to accelerated growth of planetesimals via pebble accretion \citep{Ormel2010, Lambrechts2012}. Through this mechanism planetary seeds could perhaps grow to sizes where they can survive high velocity collisions, while the cavity edge prevents them from migrating further in the disc \citep[e.g.][]{Meschiari2014}.

The trapping of dust in pressure maxima is well established in quiescent discs, where super-Keplerian gas velocity in regions with positive pressure gradient and sub-Keplerian gas velocity in regions with negative pressure gradients drives dust towards the pressure maximum \citep{Whipple1972, Weidenschilling1977, Haghighipour2003}. Since giant planets open up deep gaps in protoplanetary discs, they create pressure maxima at the outer edge of the gap that can trap dust migrating from the outer disc \citep{Paardekooper2004, Rice2006}. This may explain the origin of transition discs with large cavities \citep[e.g.][]{Pinilla2012, Zhu2012}, and has been invoked to explain the relative lack of refractory materials in photospheres of young Herbig stars with transitional discs \citep{Folsom2012, Kama2015}. In fact, the trapping of dust is possible down to masses as low as 20 to 30 $M_\oplus$ \citep{Lambrechts2014, Rosotti2016, Bitsch2018, Ataiee2018}.

However, the extension of dust trapping in pressure maxima from the planetary regime to binaries of mass ratio about $q \sim 0.1$ is not guaranteed. In the case of planets on circular orbits, the gas remains on orbits that are close to circular, while at high mass ratio the binary accretes through strong spiral streams \citep{Artymowicz1996, Rozyczka1997, Bate1997, Gunther2002}. In this paper, we demonstrate that these streams can under certain circumstances penetrate into the disc far enough to carry dust with them, thereby disrupting the trapping of dust. We study both circular and eccentric binary systems. The latter are important because circumbinary planets are often found around eccentric binaries and the behaviour of dust and gas transport in these systems needs to be understood.

Dust trapping for eccentric orbits in the giant planet regime ($q \lesssim 0.01$) is also of interest, particularly in the context of pre-transition discs \citep{Espaillat2010}, where the presence of a warm inner disc suggests that dust is transported across the gap. Eccentric planets may provide a natural explanation for the wide gaps observed \citep[e.g.][]{Cazzoletti2017}, with very massive objects or multiple planets otherwise required. We show that the timescale over which the gap opens and dust accretion stops is longer when the planetary companion is eccentric and discuss the implications of this result for pre-transition discs as well as planet formation interior to the companion's orbit.

The outline of the paper is as follows. In Section~\ref{sec:setup}, we describe the set-up of our simulations along with the physics included in our study and the important quantities we choose to quantify and study. Thereafter, we present the results of our simulations in Section~\ref{sec:dust_acc} and discuss the implications of our findings in Section~\ref{sec:discussion}. Finally, we summarise our work and suggest directions for future work in Section~\ref{sec:conclusions}.

\section{Simulation Set-up} \label{sec:setup}
We simulate the dynamics of gas and dust in the circumbinary and both circumstellar discs using Smoothed Particle Hydrodynamics (SPH), using GADGET-2 modified to include dust species \citep{Springel2005, Booth2015, Booth2016}. The equations of motion for dust and gas (respectively) are given by:

\begin{equation}
\frac{d \mathbf{v_d}}{dt} = - \frac{(\mathbf{v_d} - \mathbf{v_g})}{t_s} - \nabla \Phi(\mathbf{r})
\end{equation}

\begin{equation}
\frac{d \mathbf{v_g}}{dt} = - \frac{\nabla P}{\rho_g} - \nabla \Phi(\mathbf{r}) + \mathbf{a}_{\rm visc}
\end{equation}

Here, dust and gas particles have masses $m_d$ and $m_g$, densities $\rho_d$ and $\rho_g$, and local velocities $\mathbf{v_d}$ and $\mathbf{v_g}$ respectively. $P$ is gas pressure, $\Phi(\vec{r})$ is the gravitational potential of the binary (see appendix B) and $\mathbf{a}_{\rm visc}$ is acceleration due to artificial viscosity \citep[see][]{Springel2005}. In our simulations we compute the drag coefficient assuming constant Stokes number, $St = t_s \Omega(r)$, where $\Omega(r) = \sqrt{GM/r^3}$, the Keplerian angular frequency due to the total mass of the binary, $M$. 

The binary orbits in an anticlockwise direction and has a mass ratio $q = M_2/M_1$, total mass $M = M_1 + M_2$, semi-major axis $a$, and eccentricity $e$. The simulations are run in the inertial frame centred on the binary's centre of mass. The stars are treated as sinks and particles that come within $0.01 a$ of either star are accreted onto it \citep{Bate1995}. The binary parameters are assumed to be constant over the duration of the simulation and the stellar gravitational softening length is set to $0.001 a$, chosen to be small enough to ensure that the circumstellar discs are not affected by the softening. We do not include the effects of disc self-gravity or the back-reaction of the disc on the binary in our simulations. A globally isothermal equation of state for the gas is used with a dimensionless sound speed $c_s = 0.05$ normalised to $\sqrt{GM/a}$ (corresponding to a disc aspect ratio $H/R = 0.05 (R/a)^{1/2}$).

We use one set of particles to simulate the gas, and a second set to simulate the dust, coupled by drag forces. We work in the test-particle limit, where the influence of the dust on the gas is neglected. A quintic spline is used, with the number of neighbours set according to $\eta = 1.2$ \citep[as defined by][]{Price2012}, and the gradients involved in the pressure and artificial viscosity terms are computed using the integral-gradient SPH formulation as in \citet{Booth2016} \citep[see also][]{Rosswog2015}. A linear artificial viscosity parameter $\alpha_{SPH} = 1.5$ is employed. For our chosen parameters, this gives $\nu = \alpha_{SPH} \; c_s \; h \;/ \; 8 \approx 9 \times 10^{-3} \; h \; a \; \Omega_{\rm b}$, where $h$ is the smoothing length and $\Omega_{\rm b}$ is the binary's orbital frequency \citep{Lodato2004}. In the circumstellar discs, $h \sim 0.005 - 0.03 \; a$, which gives $\nu \sim 4 \times 10^{-5} - 3 \times 10^{-4} \; a^2 \; \Omega_{\rm b}$. In the circumbinary disc, $h \sim 0.05 \; a$, which gives $\nu \sim 4 \times 10^{-4} \; a^2 \; \Omega_{\rm b}$.

We follow \citet{Bate1997} and \citet{Young2015} and model the accretion onto the binary by injecting particles at a large distance from the binary. We define the dimensionless quantity, $R_{\mathrm{inj}}$, as the radius where the particles are injected in units of the binary semi-major axis. The particles are injected with a specific initial angular momentum $j$ that is $j_{\rm inj}$ times the binary's specific angular momentum ($\sqrt{G M a (1 - e^2)}$ ). 

We inject the particle in such a way that an accretion disc builds up over time in the simulations. To this end, we choose $R_{\mathrm{inj}} = 10$ for all our simulations apart from those with $e = 0.3$, for which we choose a larger value of $R_{\mathrm{inj}} = 20$. The velocity of the particles is set to ensure that the circularisation radius, $R_{\mathrm{circ}} = j_{\rm inj}^2 \; a \; (1-e^2)$ (i.e. the radius that the gas particles would orbit at once their orbits have circularized, assuming no exchange of angular momentum), is large enough that the gas circularises before falling onto the binary. However, since accretion rates onto the binary fall with increasing $R_{\mathrm{circ}}$ we choose $j_{\rm inj}$ to be as small as possible to best resolve the accretion \citep{Bate1997}. The injection angular momentum $j_{\rm inj}$ not only affects the circularisation radius and accretion rate but also the subsequent evolution of the location of the pressure maximum. For low mass binaries ($q = 0.01$), we inject particles with the same injection angular momentum relative to the binary for all eccentricities. This allows us to compare the relative timescales of evolution and dust trapping as a function of eccentricity alone. For higher mass ratios, we find it necessary to increase the injection angular momentum when the binary eccentricity is increased in order to form a circumbinary disc.

\begin{table}
	\centering
	\caption{Simulations Parameters}
	\begin{tabular}{cp{3cm}cc}
		\hline \hline
		Parameter & \multicolumn{2}{c}{Description} & Value \\ \hline
		$M$ & \multicolumn{2}{l}{Total Mass} & 1.0 \\
		$a$ & \multicolumn{2}{l}{Binary semi-major axis} & 1.0 \\
		$\Omega_{\rm b}$ & \multicolumn{2}{l}{Binary orbital frequency} & 1.0 \\
		$c_s$ & \multicolumn{2}{l}{Sound speed (normalised to $\sqrt{GM/a}$)} & 0.05 \\
		\multirow{2}{*}{$R_{inj}$} & \multirow{2}{3cm}{Gas injection radius (in units of a)} & $e < 0.3$ & 10 \\
		  &  & $e = 0.3$ & 20 \\
		\multirow{2}{*}{$R_{inj,dust}$} & \multirow{2}{3cm}{Dust injection radius (in units of a)} & $e < 0.3$ & 7 \\
		  &  & $e = 0.3$ & 10 \\		  
		$St$ & \multicolumn{2}{l}{Stokes number} & 0.1 \\
		$\dot{N}$ & \multicolumn{2}{l}{Gas injection rate (per $t_{dyn} = 1$)} & 500 \\
		$\dot{N}_{dust}$ & \multicolumn{2}{l}{Dust injection rate (per $t_{dyn} = 1$)} & 100 \\
        $\alpha_{SPH}$ & \multicolumn{2}{l}{Artificial viscosity parameter} & 1.5 \\
        \multirow{6}{*}{$j_{inj}$} & \multirow{6}{3cm}{Injection angular momentum for gas} & nominal & 1.2 \\
          &  & $e = 0.0, \; q =0.4$ & 1.3 \\
		  &  & $e = 0.2, \; q =0.1$ & 1.25 \\
		  &  & $e = 0.2, \; q =0.4, 0.7$ & 1.4 \\
		  &  & $e = 0.3, \; q =0.1$ & 1.4 \\
		  &  & $e = 0.3, \; q =0.4, 0.7$ & 1.5 \\
		\hline
	\end{tabular}
	\label{table:sim_param}
\end{table}

Dust particles are injected at $R_{\mathrm{inj,dust}} = 7$ ($=10$ for $e=0.3$ simulations) after 50 $t_{\mathrm{dyn}}$ ($t_{\rm dyn} = \Omega_{\rm b}^{-1}$) to allow time for the gaseous circumbinary disc to build up. A smaller dust injection radius is chosen to ensure that the dust particles have well defined SPH neighbours when they are injected, which are needed to compute the drag force. We study moderately coupled dust particles, fixing the Stokes number of the particles, $St = t_s \; \Omega = 0.1$. Here, $t_s$ is the stopping time and $\Omega$ is the Keplerian orbital frequency at a particle's location. For this choice of $St$, the angular momentum with which the dust particles are injected is unimportant because gas drag quickly brings the particles into rough corotation with the gas. Thus we inject the dust with only a slightly larger angular momentum than for the gas ($j_{\rm inj, dust} = j_{\rm inj} + 0.3$), simply because it is injected closer in than the latter (see Table~\ref{table:sim_param}).

Our choice of Stokes number is motivated by theoretical and observational constraints on protoplanetary discs, which suggest that the maximum grain size is likely in the regime {$St \approx 0.01$ -- 0.1} \citep[e.g.][]{Birnstiel2010,Tazzari2016}. Since most of the mass is typically contained in particles close to the largest size, the flux of dust arriving from the outer edge is also typically dominated by the largest particles \citep{Birnstiel2012}, hence $St = 0.1$ represents a reasonable choice when exploring the mass flux from a circumbinary disc. However, we note that inside the water snow line, $\sim 1\,{\rm au}$, dust grains are expected to be much smaller and very well coupled to the gas. In this case, results from studies of gas accretion onto binaries are expected to also apply to the dust. For reference, we note that $St=0.1$ would typically correspond to sizes of a few cm at a few au, although due to the large density variations in our simulations the exact correspondence is sensitive to location of the dust, as well as the assumed binary semi-major axis.

To quantify the delivery of material to the circumstellar discs, we follow the change in mass of both the stars and circumstellar discs, giving the net flux across their Roche lobes:
\begin{equation}
{\rm Accretion \ Rate }= \frac{\mathrm{d} (M_{disc} + M_{star})}{\mathrm{d} t}.
\label{eq:acc_rate}
\end{equation}
The circumstellar disc masses are estimated by considering the mass within the Hill radius that is bound to the star. We convolve with a Gaussian kernel of width $10 \; t_{\mathrm{dyn}}$ to remove high frequency noise in $(M_{disc} + M_{star})$ and calculate its gradient with time to determine the accretion rate. Each simulation is allowed to run for either $\sim 300 t_{\mathrm{orb}}$ or until dust accretion onto the binary ceases. We obtain steady-state accretion rates from roughly $30 t_{\mathrm{orb}} \sim 200 t_{\mathrm{dyn}}$ onwards and only include the results obtained after steady state has been established \citep[as in][]{Young2015}. To compare the delivery of dust and gas to the primary and the secondary, we use the fractional accretion rate onto the primary, $\lambda$,
\begin{equation}
\lambda = \frac{\dot{M}_{\mathrm{primary}}}{\dot{M}_{\mathrm{primary}} + \dot{M}_{\mathrm{secondary}}}.
\end{equation}
We also use the ratio $\dot{M}_\mathrm{D} / \dot{M}_\mathrm{G}$ to study the relative variation of dust and gas accretion rates with time and binary mass ratio.
\begin{equation}
\frac{\dot{M}_\mathrm{D}}{\dot{M}_\mathrm{G}} = \frac{\dot{M}_{\mathrm{primary, dust}} + \dot{M}_{\mathrm{secondary, dust}}} {\dot{M}_{\mathrm{primary, gas}} + \dot{M}_{\mathrm{secondary, gas}}},
\label{eq:dust_gas_ratio}
\end{equation}
where each $\dot{M}$ is normalized to the injection rate of the particles.

\begin{figure*}
    \centering
    \subfigure[e=0, q=0.01, 50 orbits] {\includegraphics
    [width=0.45\textwidth]{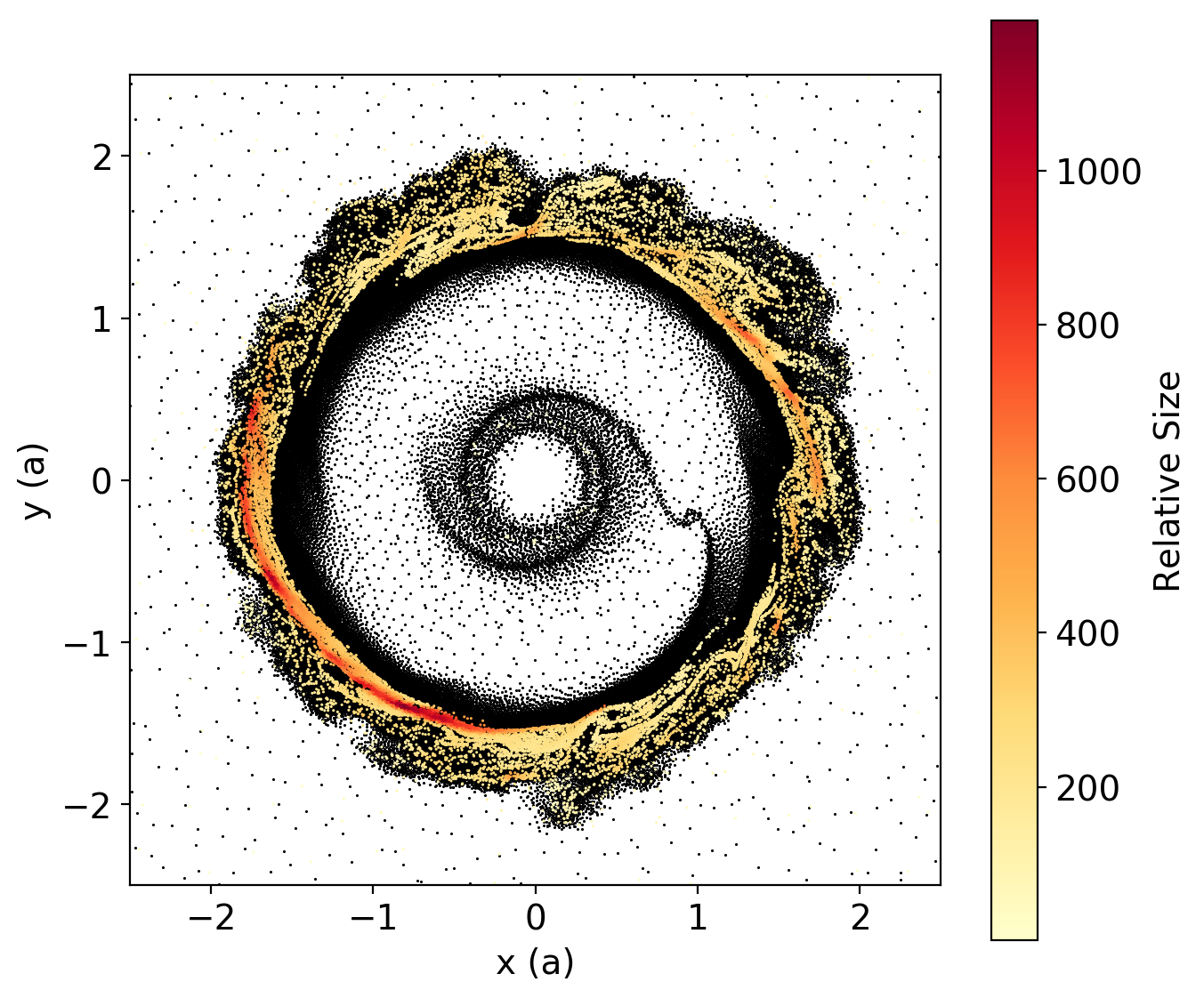}}\qquad
    \subfigure[e=0, q=0.01, 100 orbits] {\includegraphics
    [width=0.45\textwidth]{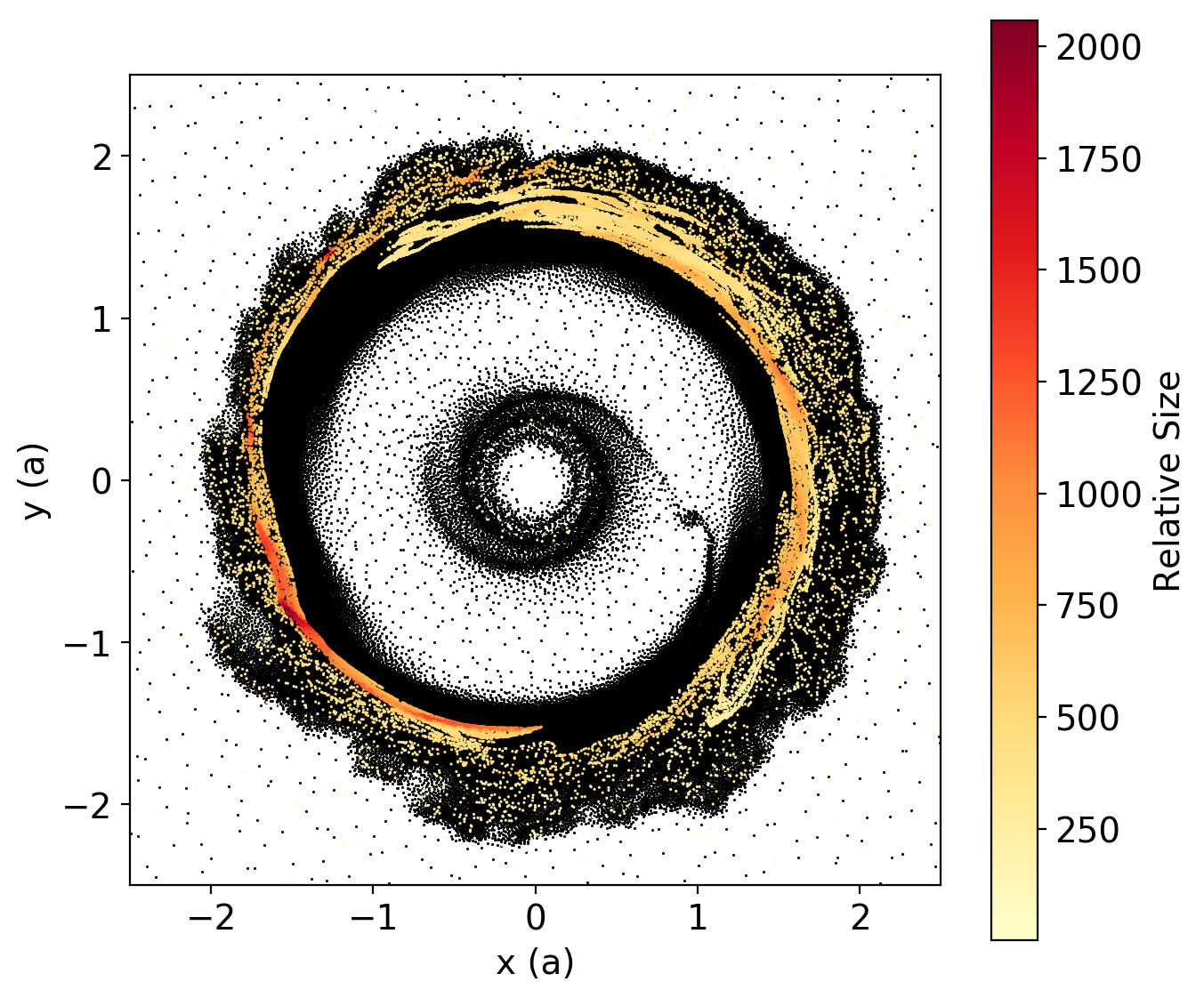}}
    \subfigure[e=0.1, q=0.01, 100 orbits] {\includegraphics
    [width=0.45\textwidth]{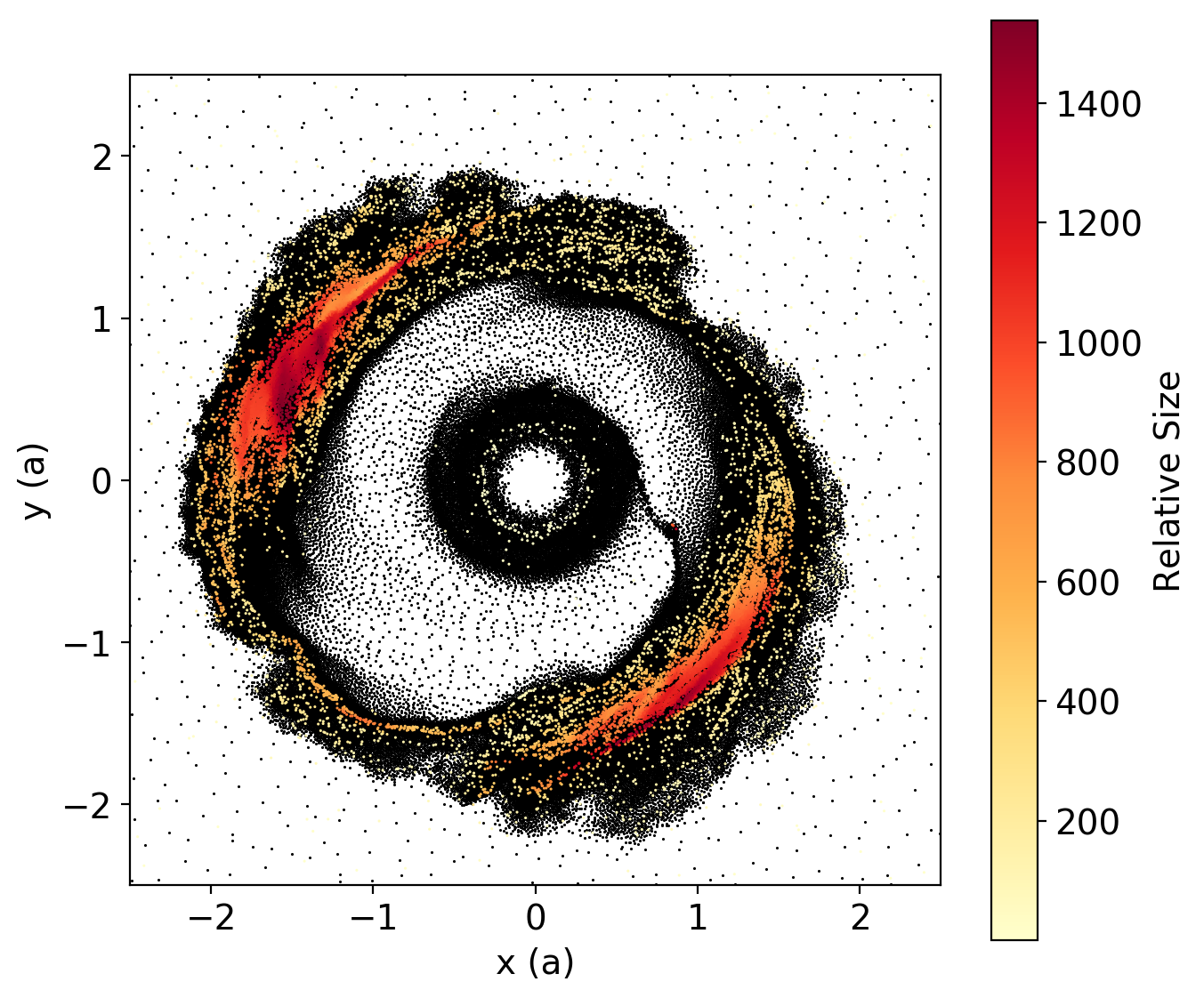}}
    \subfigure[e=0.1, q=0.01, 200 orbits] {\includegraphics
    [width=0.45\textwidth]{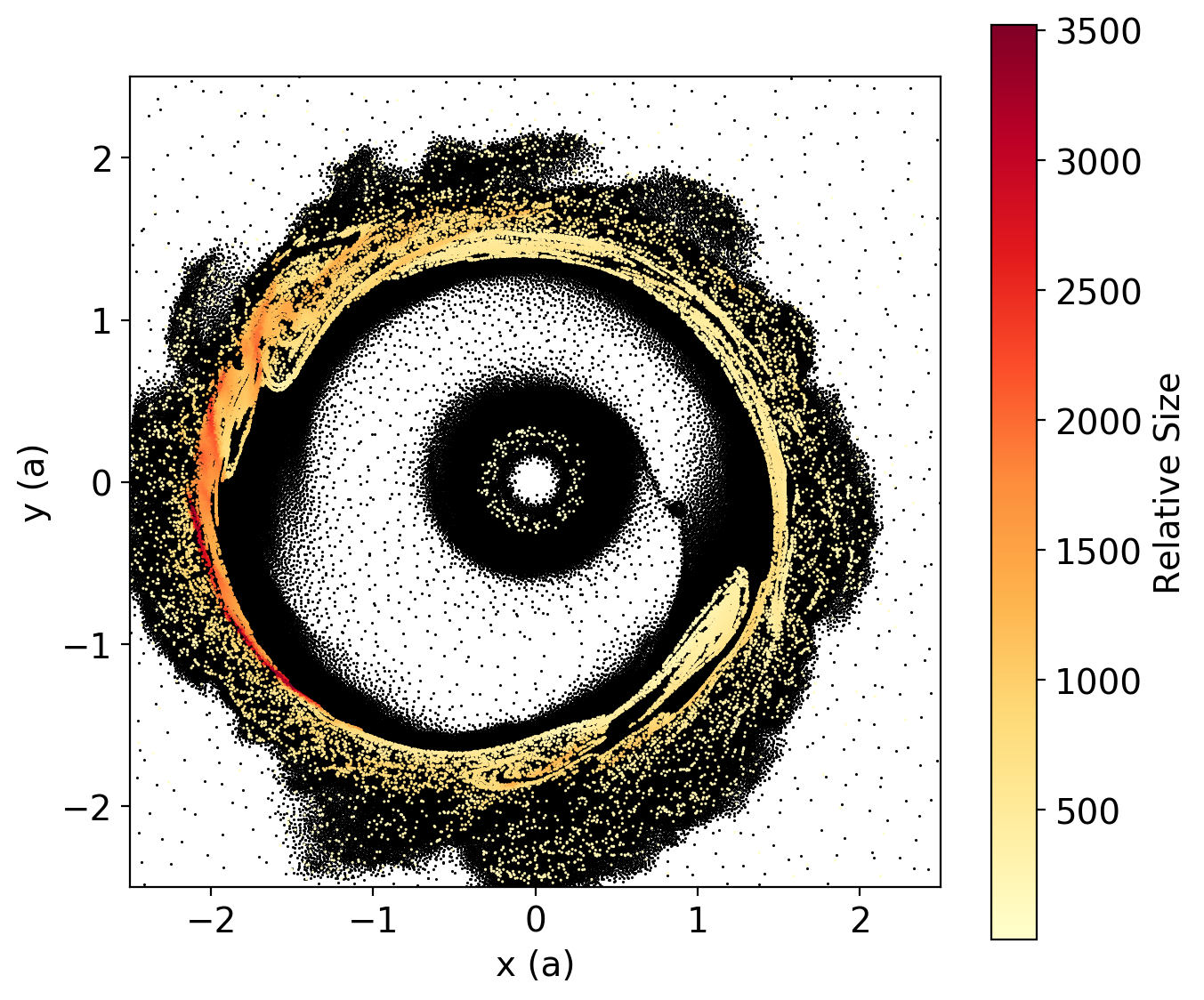}}
    \subfigure[e=0.2, q=0.01, 100 orbits] {\includegraphics
    [width=0.45\textwidth]{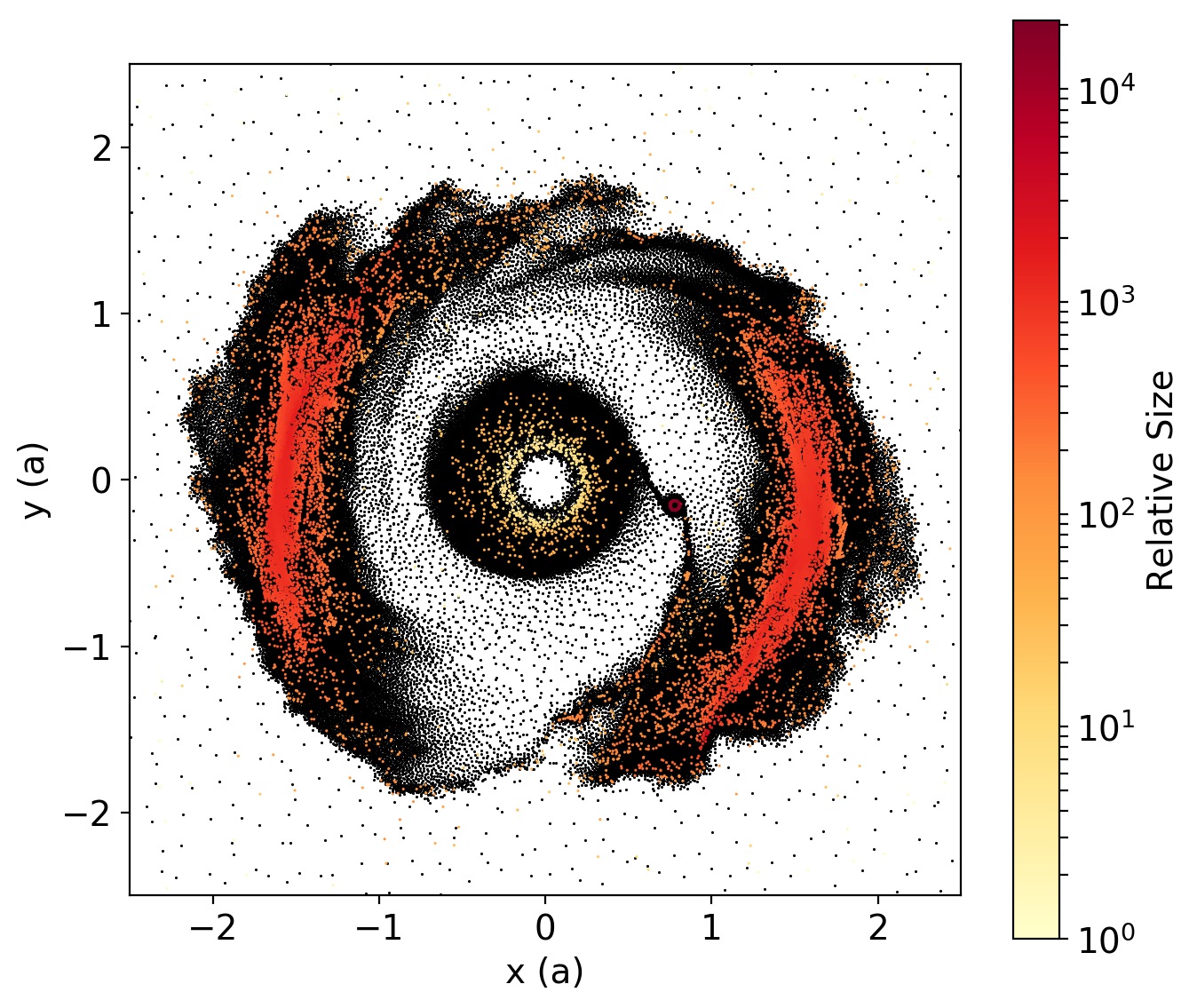}}
    \subfigure[e=0.2, q=0.01, 350 orbits] {\includegraphics
    [width=0.45\textwidth]{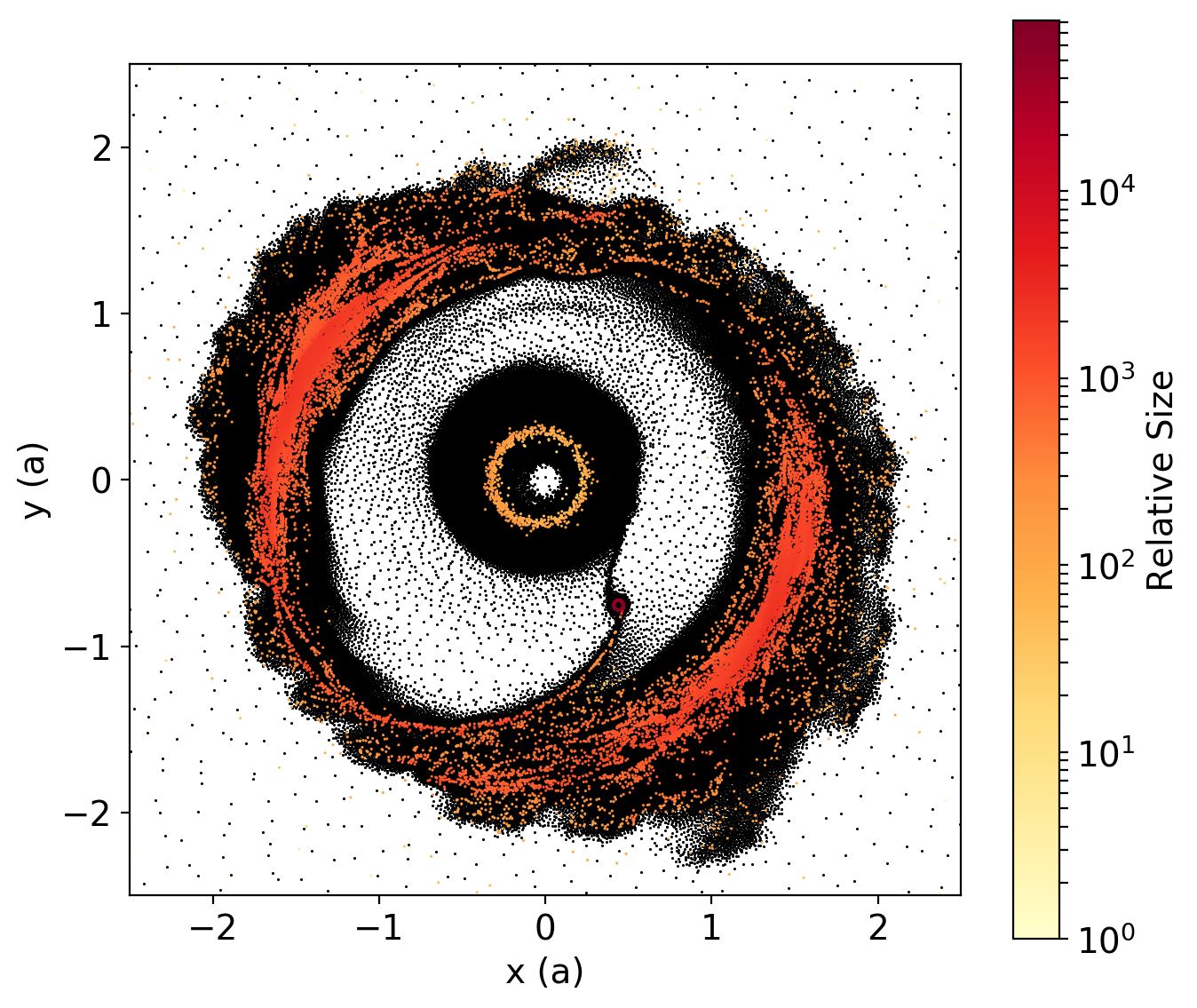}}
    \caption{Gas and dust distributions (black and coloured yellow to red respectively) for a q = 0.01 binary with different eccentricities. Dust particle sizes are calculated in the Epstein regime and normalised to the smallest size in the simulation domain; dust grains are coloured according to this relative size. 
    (a, b) Just after 50 binary orbits, accretion of $St = 0.1$ dust has completely halted for the circular binary. For $e = 0.1$ in (c) and (d), the dust accretion continues for longer and stops later than it does for the $e = 0$ case. Dust accretion continues for much longer for $e = 0.2$ binary ($>300$ orbits, panels e and f).}
	\label{fig:q01}
\end{figure*}

\begin{figure}
	\centering
  \includegraphics[width=\columnwidth]{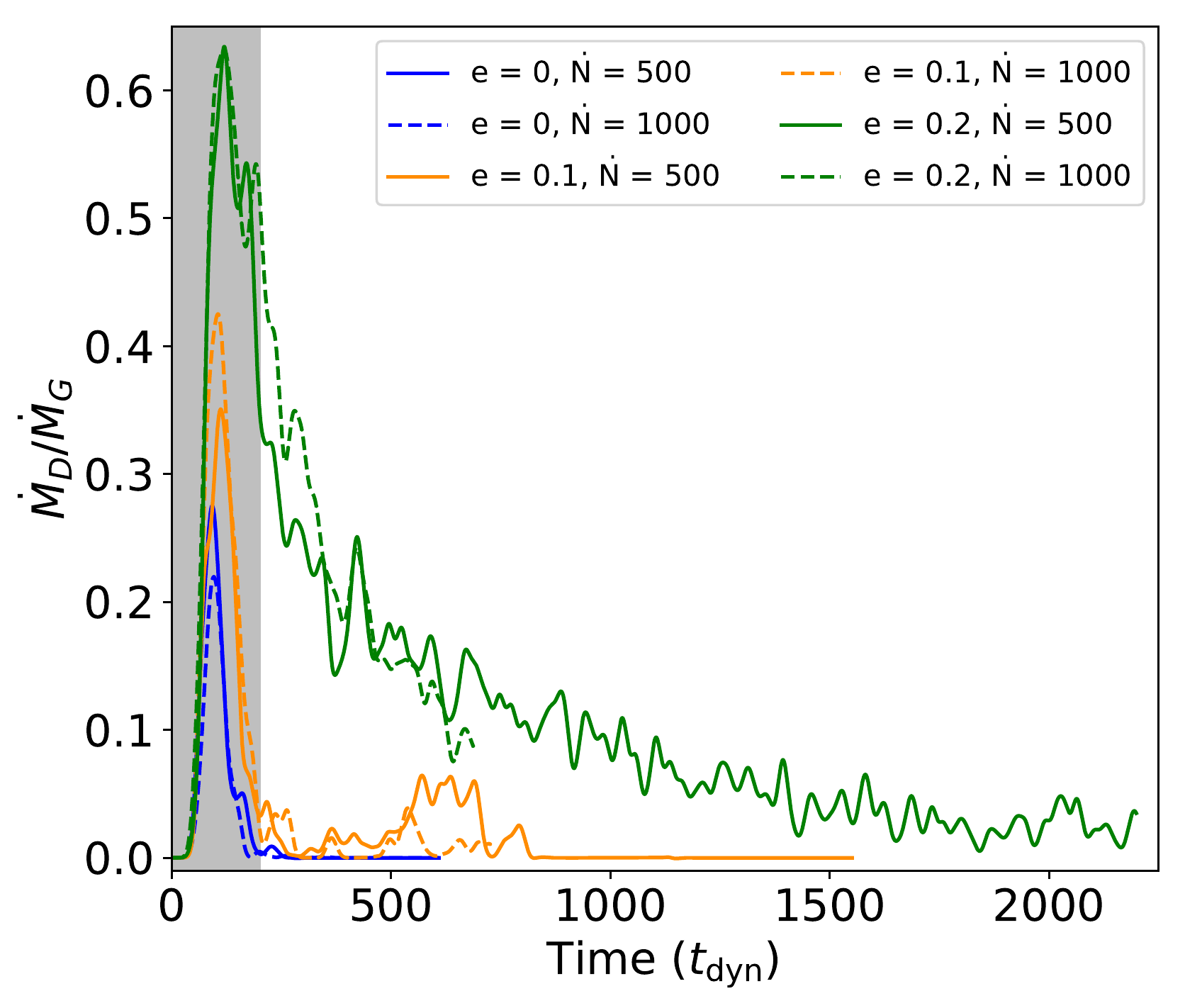}
	\caption{The ratio of accretion rates of dust and gas onto a $q=0.01$ binary (each normalised to their respective injection rates) for different eccentricities and resolutions. The initial transient phase during which the circumstellar and circumbinary discs are forming is greyed out.  After this phase, dust accretion for the circular binary and $e = 0.1$ binary is already negligible. As the eccentricity is increased, the duration for which dust accretion persists becomes much longer ($\sim 30$ orbits for a circular binary and $>300$ orbits for a binary with $e = 0.2$). Note that this plot does not show the ratio of absolute accretion rates; the y-axis needs to be multiplied by the dust-to-gas ratio in the disc to obtain absolute values.}
	\label{fig:q01_ratio}
\end{figure}

\section{Dust accretion}
\label{sec:dust_acc}

\begin{figure}
	\centering
	\includegraphics[width=\linewidth]{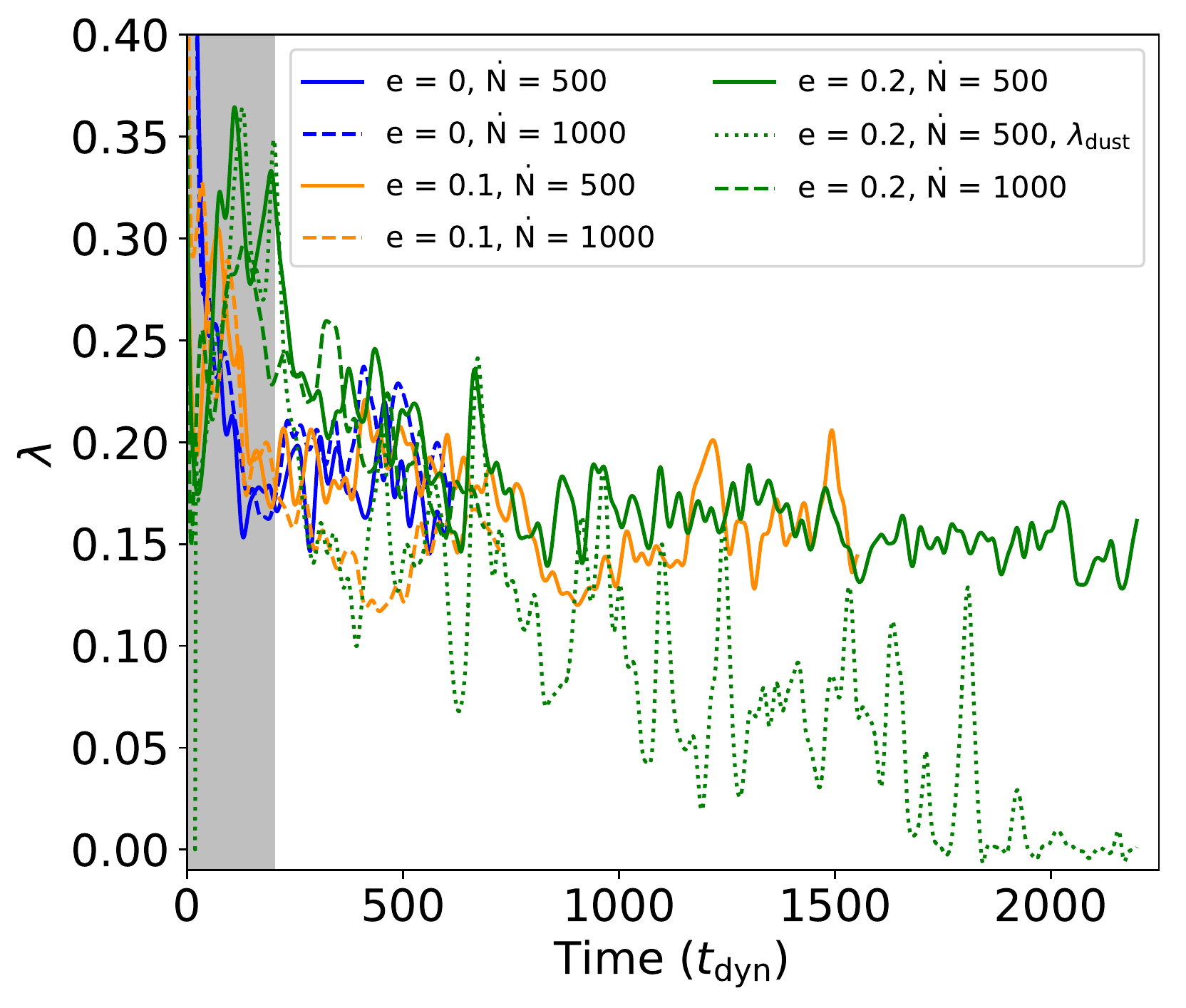}
	\caption{The fraction of gas and dust (for $e=0.2$) accreted onto the primary, $\lambda$, plotted as a function of time. After an initial transient phase ($\sim 200 t_{\mathrm{dyn}}$, greyed out), $\lambda$ for gas settles to a value of $\sim 0.15$. $\lambda$ for dust can only be properly quantified for $e = 0.2$ as dust accretion rate is too low for for lower eccentricites. There is a clear decline in $\lambda$ for dust, implying that the amount of dust accreted onto the binary (Figure~\ref{fig:q01_ratio}) as well as onto the primary is going down with time.}
	\label{fig:lambda_q_01}
\end{figure}

\subsection{Planetary regime} \label{sec:q001}
We begin by considering a binary with mass ratio $q=0.01$ (i.e. a $10\,M_{\rm J}$ planet for a solar mass star). Here, it is well known that companions on circular orbits are able to trap dust in pressure maxima that form outside of their orbit, as is suspected to be the origin of transition discs \citep[e.g.][]{Pinilla2012,Zhu2012,Kama2015}. From \autoref{fig:q01}, it is clear that dust particles (with $St = 0.1$) are trapped efficiently for companions on a circular orbit. Just after $\sim 30$ binary orbits, the accretion onto the secondary is purely gaseous (black particles represent gas) and the accretion stream does not reach the pressure maximum in which the dust is trapped. However, it is evident that the binary takes a significantly longer time to reach this stage in the case of non-zero eccentricity (for the same injection angular momentum relative to the binary). For $e = 0.1$, there is marginal dust accretion onto the binary initially ($\lesssim 100 t_{\mathrm{orb}}$) but by the time 200 binary orbits have elapsed, dust is entirely trapped in the circumbinary disc and the accretion streams are purely gaseous. For $e = 0.2$, the timescale is even longer, with ongoing dust accretion even after 300 binary orbits.

To quantify the accretion of dust, we plot the ratio of the dust accretion rate to the gas accretion rate onto the binary in \autoref{fig:q01_ratio}. This plot quantitatively reinforces the points made above and demonstrates that our conclusions are not a result of the resolution of our simulations. The decline in dust accretion for $e=0.2$ simulation is gradual and we expect dust accretion to stop eventually. For higher eccentricities, dust accretion would probably continue for even longer. Notably, Figure~\ref{fig:q01_ratio} shows the accretion of dust onto both the primary and the secondary. We show the fraction of dust and gas accreted onto the primary ($\lambda$) in Figure~\ref{fig:lambda_q_01} to compare the relative distribution of accreted circumbinary material onto the two components of the binary.

For the three eccentricites considered here, $\lambda$ for gas takes a roughly constant value of 0.15 once the simulations reach pseudo-equilibrium, implying that nearly 15\% of the gas accreted from the circumbinary disc goes to the primary. We do not observe a strong dependence of $\lambda$ for gas on the binary eccentricity. $\lambda$ for dust is qualitatively different in that it declines with time. For $e = 0$ and $e = 0.1$, the dust accretion phase is comparable to the time it takes for the simulations to reach pseudo-equilibrium. Therefore, it is not possible to get a reasonable estimate of $\lambda$ for these simulations. For $e = 0.2$, we plot $\lambda$ for dust as a function of time and find that it declines with time. This implies that the rate at which dust is accreted by the primary is declining more rapidly than the corresponding rate for the binary (both primary and secondary).

\subsection{Stellar regime} \label{sec:q01_1}
\begin{figure}
	\centering
	\includegraphics[width=\linewidth]{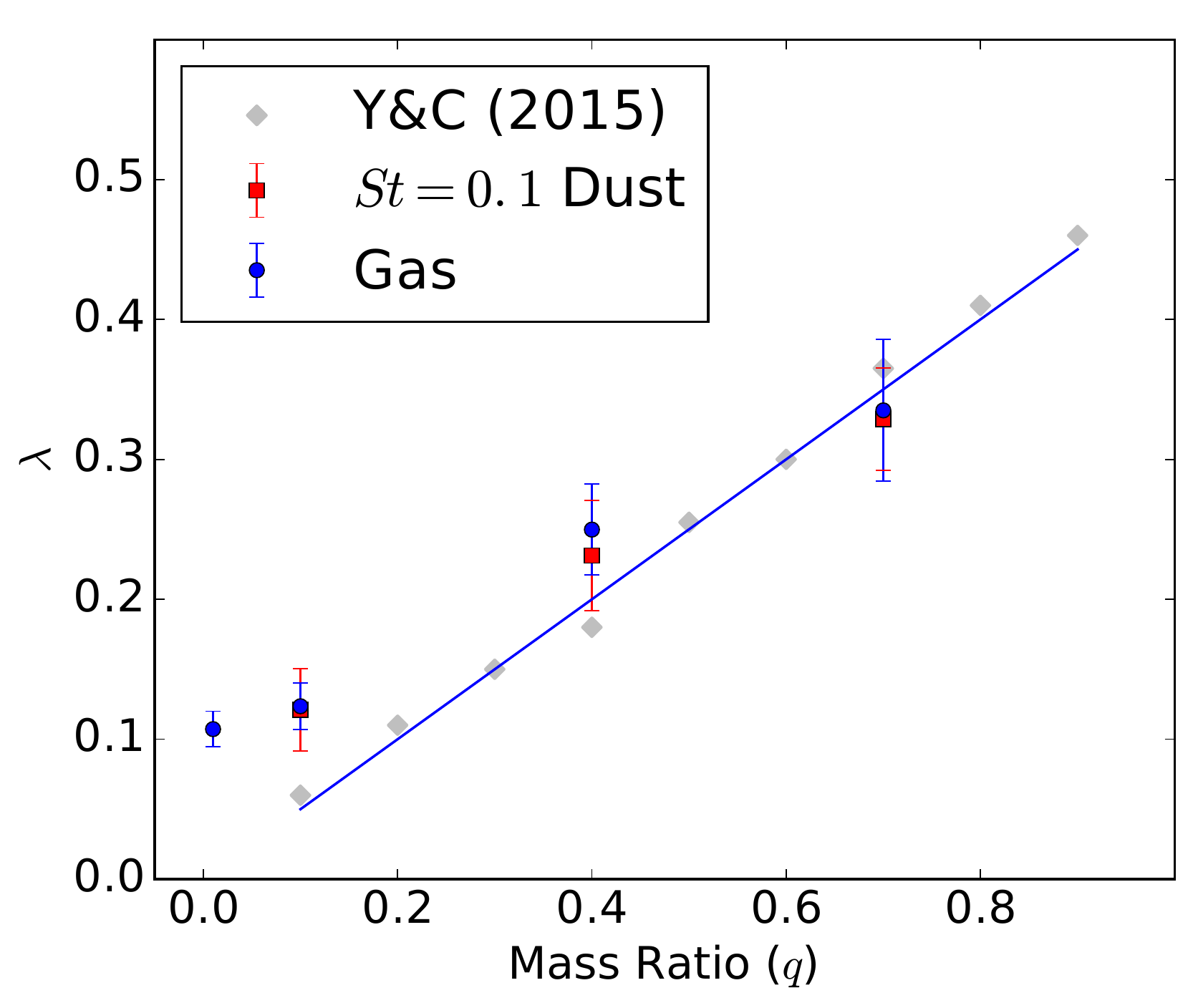}
	\caption{The fraction of gas and dust accreted onto the primary, $\lambda$, for a circular binary as a function of mass ratio. The error bars show the scatter in the distribution of $\lambda$. For $q = 0.01$, a higher resolution simulation ($\dot{N} = 2000$) is used for this plot and $\lambda$ for dust is undefined as there is no dust accretion. For $q \gtrsim 0.1$, $St=0.1$ dust and gas show the same fractional division between secondary and primary. Our results also compare well with the results (grey dots and $\lambda = q/2$ parametrisation) obtained by \citealp{Young2015a}.}
	\label{fig:dust_gas_acc}
\end{figure}

In the case of binaries where the companion is in the stellar regime ($q \gtrsim 0.1$), we find that there is initially a flow of dust onto the circumstellar discs. During this dust accretion phase, we find that $St=0.1$ dust is divided between the primary and the secondary in the same ratio as the gas and thus the dust to gas ratio in each of the circumstellar discs is similar (\autoref{fig:dust_gas_acc}). This can be understood inasmuch as pressure only plays a sub-dominant role in the accretion of gas at aspect ratios $H/R \sim 0.05$, being more important at higher temperatures \citep{Young2015}. Since the differences between the dynamics of dust and gas are already small for $St=0.1$, we expect this result to hold for smaller particles as well since they are more strongly coupled to the gas. We also compare $\lambda$ from our simulations with values obtained by \cite{Young2015a} and show that there is good agreement between the two in \autoref{fig:dust_gas_acc}.

\begin{figure*}
	\centering
	\includegraphics[width=\textwidth]{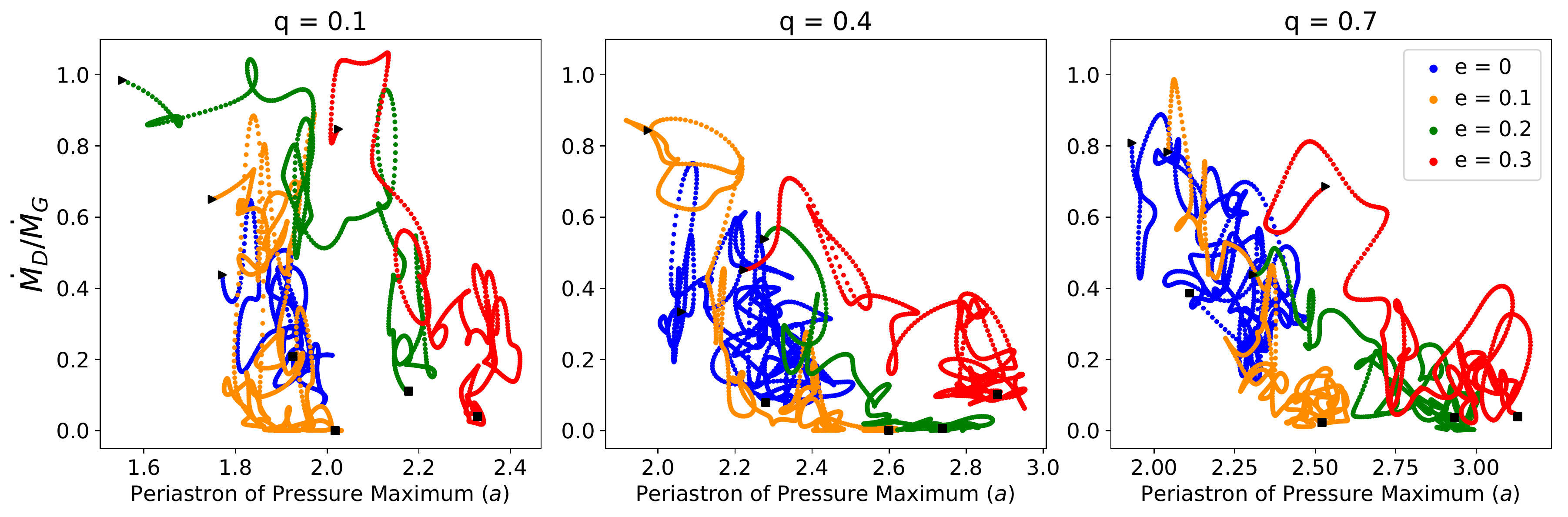}
	\caption{Plots showing the ratio of dust accretion to gas accretion against the mean periastron of the pressure maximum. The accretion rate ratio is calculated by taking the gradient of the total accreted mass (star + disc) convolved with a Gaussian filter 10 $t_{dyn}$ wide (Equation~\ref{eq:dust_gas_ratio}). We calculate the periastron of the pressure maximum by finding the mode of the semi-major axis distribution of the circumbinary disc and thereafter calculating the mode of periastron of these fluid elements. The ability of the binary to accrete dust from the circumbinary disc's pressure maximum depends on the maximum's position. The time evolution begins with with $\blacktriangleright$ and ends with $\blacksquare$. The outward movement of the disc periastron correlates with decline in dust accretion.}
	\label{fig:per_peak_var}
\end{figure*}

However, as the disc evolves in time and the gap opened by the binary becomes wider, dust accretion declines considerably and stops in most of our simulations (\autoref{fig:per_peak_var}). To investigate the origin of this behaviour, we considered the hypothesis that dust could be trapped in a pressure maximum away from the inner edge of the disc, while the binary is able to accrete gas from interior to this maximum. With this in mind we tested whether the periastron radius of fluid elements located in the pressure maximum provided a good predictor of whether the binary was able to accrete dust. Figure~\ref{fig:per_peak_var} shows the evolution of the ratio of dust to gas accretion rates (where a value of unity would indicate that the binary accretes dust and gas in the same ratio as the input dust to gas ratio) versus the pericentre of pressure maximum for different mass ratios and eccentricities. It illustrates that the accretion of dust stops once the periastron distance exceeds the binary separation by a factor in the range $\sim 2.2-2.8$. {\footnote {Note that we have also investigated how dust accretion instead correlates with the semi-major axis of fluid elements in the pressure maximum but find that this quantity less cleanly delineates the regime where dust can accrete.}} Moreover, at higher $q$ there is a trend for the pressure maximum to need to reach larger distances before dust accretion is inhibited.

The location of the pericentre radius increases in time, i.e. the gap opened by the binary continues to widen. However, the details of this evolution depend on both the properties of the binary, and in part on the viscosity and the angular momentum with which the gas particles are injected. Given that we run the simulations for a relatively short period of time, the discs do not reach a steady state and the gaps are continuing to open. We find that this time dependence of the pericentre radius helps explain why models transition between a state of dust accretion and dust trapping. Since the simulations of \citet{Thun2017} are however evolved to a steady state, we can use their results to infer whether the accretion of dust or trapping is expected more generally. \citet{Thun2017} find that radius of the pressure maximum is typically between 3 to 5 binary separations, although this value is decreased for larger viscosity and higher aspect ratios. Based on our finding that dust accretion stops once the dust pericentre recedes beyond 2.2-2.8 times the binary semi-major axis, together with the dependence of the cavity radius on viscosity determined by \citet{Thun2017}, we suggest that it is unlikely that accretion of dust with $St=0.1$ should be able to persist once the circumbinary disc has evolved into a steady state, unless $\alpha h^2 \gtrsim 10^{-4}$. Here, $\alpha$ is the alpha viscosity parameter \citep{Shakura1973} and $h$ is the disc aspect ratio.

For dust grains with smaller $St$ ($< 0.1$) we expect the differences between the dust and gas to be smaller. This is because although dust grains inside the pressure maximum drift outwards towards the pressure maximum, they are also replaced by diffusion. Since turbulent diffusion is not sensitive to Stokes number \citep[e.g.][]{Youdin2007}, but radial drift is slower for smaller $St$, this will result in higher dust surface densities inside the pressure maximum, and thus more efficient accretion of dust at larger cavity sizes. At even smaller St values, the drift velocity must become smaller than the gas radial velocity, resulting in the gas and dust accreting at approximately the same rate.

\section{Discussion}
\label{sec:discussion}

The accretion of dust in binary systems is important for the formation of planets in binary systems, with there being two distinct possibilities: 1) dust is trapped in a pressure maximum in the circumbinary disc or 2) dust passes through the circumbinary disc and enters the circumstellar discs. Alternatively in the intermediate scenario there is some accretion of dust onto the individual binary components but the resulting dust to gas ratio is low. In the first scenario, {\it in situ} formation of planets in a circumstellar configuration may be difficult due to a lack of solid material available, but the build up of solids at the inner edge of the circumbinary disc may aid the {\it in situ} growth by pebble accretion of circumbinary planets, which are known to be close to the \citet{Holman1999} stability limit \citep{Martin2014}. In the second scenario, the reverse may be true and the assembly of rocky planets may occur in circumstellar, as opposed to circumbinary, discs. Here we find evidence that both regimes may exist, depending on the binary parameters, and the size of dust grains.

For binaries where both objects are of comparable mass ($q \gtrsim 0.1$, the `stellar' regime), we find that dust accretion stops once the circumbinary disc recedes to a large enough distance from the binary. The equilibrium cavity size of discs around protostellar binaries is likely to be large enough that dust trapping can occur, which might seem favourable for {\it in situ} formation of circumbinary planets. However, we note that the cavity sizes are typically larger than the observed planet semi-major axes, and thus formation of planets at the pressure maximum would not really be {\it in situ} -- some migration would still be required to bring them to their current location.

The `planetary' mass regime, i.e $q \lesssim 0.01$, also has interesting consequences for planet formation. For circular systems, massive planets are known to open deep gaps on short timescales ($\sim 30$ orbits in our simulations), preventing the accretion of dust onto the star and possibly forming transition discs. Here we show that if the planet has significant eccentricity ($e\gtrsim 0.2$), the transient dust accretion phase lasts for a substantially longer duration ($\gtrsim 300$ orbits). Keeping in mind the caveat that this timescale is somewhat dependent on the chosen numerical parameters, we find that it is roughly comparable to the drift time to the star for $St = 0.1$ particles in an unperturbed disc but much shorter than the $\sim$ Myr timescale required for planet formation in the inner disc \citep{Armitage2010}. Extrapolating our results to higher eccentricities, it is likely that the dust accretion declines over longer timescales as the binary eccentricity is increased. Nonetheless, at $\sim 1$ AU, these timescales are short enough that they cannot significantly impact the dust budget of the circumprimary/inner disc. For anomalous systems such as WASP-53 and WASP-81A, which have massive (brown dwarf) secondaries on high eccentricity orbits (0.83 and 0.55) at small semi-major axes ($\sim 2 - 4$ AU) and yet possess hot Jupiters \citep{Triaud2017}, in situ formation in the truncated circumprimary disc \citep{Artymowicz1994, Haghighipour2008} seems highly unlikely. A more plausible solution is that the eccentric companions attained their current orbits after the planets had formed already.

Pre-transitional discs provide an interesting test case for dust transport across gaps in discs given the presence of both wide gaps in the millimetre emission and evidence for warm inner disc (e.g. LkCa 15 or UX Tau A, \citealt{Espaillat2010}). The traditional explanation for these inner discs is that small dust grains (smaller than those we consider here) may be accreted along with the gas \citep[e.g.][]{Pinilla2016}. Eccentric planets are an attractive solution to the origin of pre-transitional discs because they allow wide gaps to be opened without the need for multiple planets or very high masses \citep[Figure~\ref{fig:q01}; see also e.g.][]{Cazzoletti2017}. In Figure~\ref{fig:q01}, we see that the planetary companion opens a large gap which is devoid of $St = 0.1$ dust, which would correspond to the observed gap in pre-transitional discs. While we see that for eccentric planets the large grains considered here ($St=0.1$ mm to cm size) can be transported into the inner discs for longer times than circular planets, this time is relatively short compared with the expected lifetime of pre-transitional discs ($\sim 100$ orbits, i.e. $10^4$ years for $St = 0.1$ particles, assuming $e = 0.2$ and planet at 20 au; \citealp{Espaillat2014}). Thus it seems more likely that the inner discs seen in pre-transitional discs are due to grains smaller than those considered here being carried along with the gas, rather than due to high eccentricity alone. A more rigorous comparison of the timescale during which accretion lasts and the duration of the pre-transitional disc phase, especially for higher binary eccentricities and smaller dust grains, is warranted in the future.

An additional property of these systems that could play an important role in dust transfer from the outer to the inner disc is the inclination of the secondary relative to the disc. Even for moderate inclinations (sin$(i) > H/R$), the companion spends a large fraction of its orbit away from the disc mid-plane. Gaps opened by inclined companions are known to be shallower and narrower than those opened by co-planar companions \citep[e.g.][]{Chametla2017}. This may make it easier for both dust and gas to accrete onto the inner disc and aid planet formation in this region. Ultimately, the importance of inclination's role in dust transfer past the companion will depend on how the timescales for gap opening, inclination damping, and accretion compare with each other \citep[e.g.][]{Chametla2017, Nealon2018}. We encourage future work that assesses the role of both inclination and eccentricity in the transfer of dust past a planetary companion.

\section{Conclusions}
\label{sec:conclusions}
We have shown that the combination of dust and gas dynamics in binary systems exhibits a rich behaviour, depending on whether the binary companion is in the stellar regime or is a massive planet.

While it is well known that suitably sized dust is trapped in the outer disc in the case of a planetary mass companion on a circular orbit, we have shown that accretion of dust onto the individual binary components continues for an appreciably longer time in the case of moderately eccentric planetary orbits. Nonetheless, this timescale is much shorter than planet formation timescale in the inner disc. This makes it challenging for planets to form in situ interior to orbits of massive companions at a few AU that prevent the transfer of solids from the outer disc to the inner disc. In addition, although eccentric planets offer an interesting explanation for pre-transitional discs, accretion timescale for dust are likely to be shorter than the lifetime of this disc phase. We suggest that companion inclination may play an important role along with eccentricity in determining the fate of solids in the outer disc.

Our investigation for stellar mass ratios ($q \gtrsim$ 0.1) reveals that there is a phase of dust accretion onto the binary as the dust trap is disrupted by the binary's strong non-axisymmetric potential. During this phase, the destination (i.e. primary or secondary star) of moderately coupled dust ($St=0.1$) is a function of mass ratio and closely follows that of the gas. This phase however turns out to be a transient since the location of the pressure maximum (where dust accumulates in the circumbinary disc) undergoes secular outward migration. We find empirically that once the periastron radius of dust and gas in the pressure maximum exceeds $\sim 2.2-2.8$ times the binary separation (with some systematic increase in this radius as $q$ increases), dust remains in the trap and is no longer accreted into the discs around the individual stars. Previous studies suggest that all systems are likely to evolve to a state where the pressure maximum is sufficiently far out \citep[e.g.][]{Thun2017} that we would not expect accretion of dust in this case.

The trapping of dust outside the binary orbit could have implications for the {\it in situ} formation of circumbinary planets. The chances of {\it in situ} formation by gravitational collapse, or the rapid growth of a planetesimal that has migrated to the pressure maximum, are enhanced if the dust-to-gas ratio becomes high in this region. Further theoretical work is needed to establish whether {\it in situ} formation via gravitational or fluid instabilities is possible in the pressure maxima of circumbinary discs. Atmospheric composition studies using transmission spectroscopy of circumbinary planets would also be invaluable in distinguishing between {\it in situ} and {\it ex situ} formation mechanisms.

\section*{Acknowledgements}
We are grateful to the referee for thoughtful comments and suggestions that substantially improved this manuscript. Y. C. is grateful to St John's College, Cambridge and the Cambridge Commonwealth Trust for their generous support during the course of his education. Y. C. also thanks Eve J. Lee for helpful discussions. This work has been supported by the DISCSIM project, grant agreement 341137 funded by the European Research Council under ERC-2013-ADG. This project has received funding from the European Union's Horizon 2020 research and innovation programme under the Marie Sk\l{}odowska-Curie grant agreement No 823823 (DUSTBUSTERS). Part of this work was undertaken on the COSMOS Shared Memory system at DAMTP, University of Cambridge operated on behalf of the STFC DiRAC HPC Facility. This equipment is funded by BIS National E-infrastructure capital grant ST/J005673/1 and STFC grants ST/H008586/1, ST/K00333X/1. Additionally, this work was performed also using the DiRAC Data Intensive service at Leicester, operated by the University of Leicester IT Services, which forms part of the STFC DiRAC HPC Facility (www.dirac.ac.uk). The equipment was funded by BEIS capital funding via STFC capital grants ST/K000373/1 and ST/R002363/1 and STFC DiRAC Operations grant ST/R001014/1. DiRAC is part of the National e-Infrastructure.

\bibliographystyle{mnras}
\bibliography{mnras_template}

\appendix
\section{Resolution Study}
\begin{figure}
	\centering
	\includegraphics[width=\linewidth]{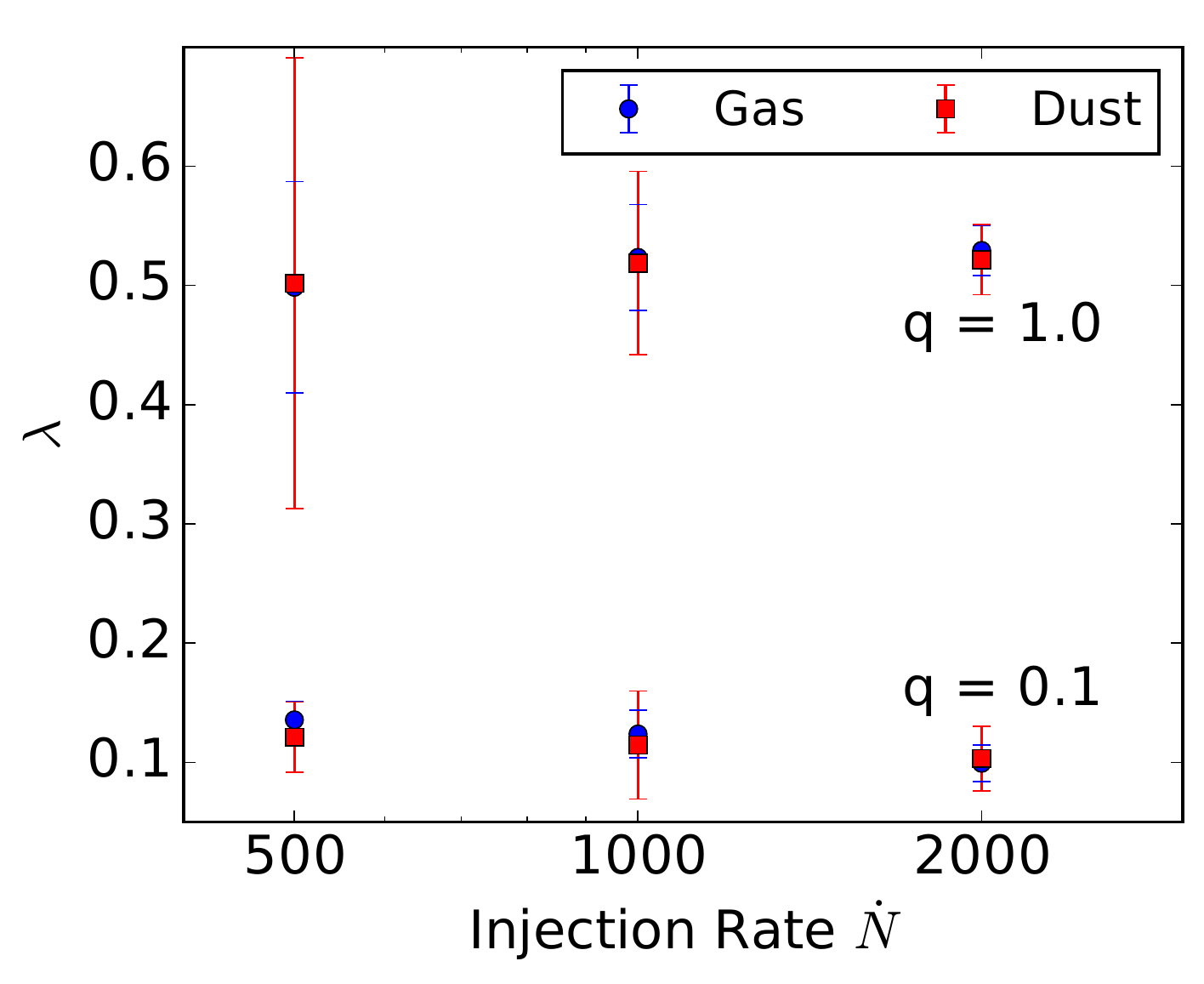}
	\caption{Resolution study for dust and gas simulations conducted for $q$ = 0.1 and $q$ = 1.0 for three different $\dot{N}_{gas}$ = 500, 1000, 2000 particles per $t_{\mathrm{dyn}}$ with $St$ = 0.1 dust particles. The estimate for $\lambda$ obtained for the three simulations are consistent with each other. The accretion rate of dust is closely coupled to that of gas in the $St$ = 0.1 regime. Additionally, as $\dot{N}$ is increased, the noise in the estimate reduces significantly. }
	\label{fig:res_study}
\end{figure}

Initially, we test our simulations by running a resolution study for a circular binary with $q$ = 0.1 and $q$ = 1.0. The particle injection rate $\dot{N}_{\mathrm{gas}}$ is increased from 500 to 1000 and 2000 particles per $t_{\mathrm{dyn}}$ (along with $\dot{N}_{\mathrm{dust}}$= 100, 200, \& 400 particles with $t_{\mathrm{dyn}}$) to check for convergence in the accretion rates. In Figure~\ref{fig:res_study}, we show that as the resolution of the simulation is increased ($\dot{N}$ is increased), $\lambda$ remain the same. The global average for the accretion rate is also consistent for three values of $\dot{N}$, indicating that the simulations are working reliably. From Figure~\ref{fig:res_study}, we see that although the noise in $\lambda$ decreases with increasing resolution, the mean value is already well reproduced at an injection rate of $\dot{N}_{\mathrm{gas}}$ = 500 particles per $t_{\mathrm{dyn}}$. We therefore use that as our base resolution ($\dot{N}_{\mathrm{gas}}$ = 500 and $\dot{N}_{\mathrm{dust}}$= 100 particles per $t_{\mathrm{dyn}}$).

\section{Gravitational acceleration}
The gravitational potential of the binary is included by directly computing the acceleration due to the binary on each particle, which has been softened on a scale of $10^{-3}a$. Here, we follow \citet{Price2007} and compute the acceleration by spreading the mass over a volume defined by the quintic spline smoothing kernel. This has the advantage that the force is exactly the Newtonian acceleration for separations greater than the softening length, $\epsilon$. 

For $r > \epsilon$, the force due to the gravity from each star is given by 
\begin{equation}
\mathbf{g} = - G M_i \frac{\mathbf{r}}{r^3}.
\end{equation}
For $r < \epsilon$, we define $s = 3 \times r/\epsilon$ and compute the softened gravity according to:
\begin{equation}
\mathbf{g} = - \frac{324}{359} \frac{G M_i \mathbf{r}}{\epsilon^3} \times \begin{cases}
 - \cfrac{15}{4}s^5 + \cfrac{90}{7} s^4 - 36 s^2 + 66 , & s < 1, \\
 
\cfrac{15}{8} s^5 - \cfrac{135}{7} s^4 + 75 s^3 - \\
126 s^2 + \cfrac{225}{4} s + 51 + \cfrac{5}{56}\cfrac{1}{s^2} , & 1 <s <2 , \\

-\cfrac{3}{8} s^5 + \cfrac{45}{7} s^4 - 45 s^3 + \\
162 s^2 - \cfrac{1215}{4} s + 243 - \cfrac{507}{56}\cfrac{1}{s^2} , & 2 < s < 3
\end{cases}
\end{equation}

\bsp
\label{lastpage}
\end{document}